# Time-resolved temperature mapping leveraging the strong thermo-optic effect in phase-change devices

Nicholas A. Nobile[1], John R. Erickson[1], Carlos Ríos[2,3], Yifei Zhang[4], Juejun Hu[4], Steven A. Vitale[5], Feng Xiong[1], Nathan Youngblood[1,*]

[1]Univ. of Pittsburgh, Dept. of Electrical & Computer Engineering, Pittsburgh, PA 15261, USA,
[2]Univ. of Maryland, Dept. of Materials Science & Engineering, College Park, MD, USA
[3]Univ. of Maryland, Institute for Research in Electronics & Applied Physics, College Park, MD, USA
[4]MIT, Dept. of Materials Science & Engineering, Cambridge, MA, USA
[5]Advanced Materials and Microsystems Group, MIT Lincoln Laboratory, Lexington, MA

*Corresponding email: nathan.youngblood@pitt.edu

**Abstract:** Optical phase-change materials are highly promising for emerging applications such as tunable metasurfaces, reconfigurable photonic circuits, and non-von Neumann computing. However, these materials typically require both high melting temperatures and fast quenching rates to reversibly switch between their crystalline and amorphous phases—a significant challenge for large-scale integration. Here, we present an experimental technique which leverages the thermo-optic effect in GST to enable both spatial and temporal thermal measurements of two common electro-thermal microheater designs currently used by the phase-change community. Our approach shows excellent agreement between experimental results and numerical simulations and provides a non-invasive method for rapid characterization of electrically programmable phase-change devices.

**Introduction:**

The field of optical phase-change materials (PCMs) has enjoyed a renaissance in the last decade since the proposal[1] and demonstration[2,3] of nonvolatile, multilevel memory integrated on photonic waveguides. Since these demonstrations, applications for optical PCMs have rapidly expanded to tunable metasurfaces[4–10], photonic computation[11–15], programmable phononic control[16,17], reconfigurable photonic circuits[18–21], plasmonic circuits[22–25], and beyond[26]. However, despite their desirable optical tunability and stability, reliable and reversible control of these materials is challenging to achieve using integrated electrical methods. This stems from stringent thermal requirements during the amorphization process in optical PCMs such as $Ge_2Sb_2Te_5$ (GST), $Ge_2Sb_2Se_4Te_1$ (GSST), $Sb_2Se_3$, and others. While crystallization temperatures range from 120 °C to 300 °C, the phase-change chalcogenides that are of interest for optical devices typically share a similar melting temperature near or above 600 °C[27]. Additionally, for PCMs with fast crystallization dynamics, such as GST, the required quenching rates are estimated to be around ~1 °C/ns to enable reamorphization[27,28]. (Note that this critical cooling rate is much lower for optical PCMs with slower crystallization dynamics, such as GSST[29].) Both these conditions are relatively simple to achieve with optical pulses in thin GST films since their significant crystalline absorption enables localized thermal annealing and rapid quenching[30,31]. However, for large-area devices with dimensions much larger than the optical wavelength (i.e., greater than ~10λ), it is non-trivial to design reliable electro-thermal switching devices which achieve uniform heating and rapid quenching profiles across the PCM using electro-thermal switching approaches.

Recently, significant progress has been made to demonstrate reversible electrical switching of optical PCM devices using resistive microheaters comprised of silicon[20,32,33], metal[10,25,34,35], transparent conductive oxides[26,36,37], and graphene[38,39]. These indirect electro-thermal approaches decouple Joule heating from the conductance of the PCM, overcoming the challenge of short circuits which plague designs where currents pass directly through the PCM. This has led to designs that vary widely in their efficiencies, speeds, and robustness on account of their distinct thermal dynamics. While electro-thermal simulations have been applied to optimize and understand the thermal response of these devices[40,41], an experimental approach which spatially maps the peak temperatures and quenching rates of these high-speed devices after fabrication is lacking. Here, we present an experimental technique which allows us to spatially map the dynamic thermal response of a GST pixel on Pt and doped-silicon microheaters during the application of electrical pulses (**Figure 1**). Our non-invasive technique makes use of the strong thermo-optic (TO) response in GST[42] to measure changes in temperature via changes in the reflection of an optical probe, providing fast and localized information on the system's thermal response.

**Experiment and Discussion Section:**

Many techniques exist by which temperature can be either spatially or temporally mapped with high resolution, but few solutions are suitable for measuring integrated optical PCM systems non-invasively while simultaneously offering necessary resolution in both temporal and spatial domains. For example, Raman thermoreflectance measurements[43] and AFM thermoreflectance[44] techniques offer sub-micron spatial resolution but are typically low speed (i.e., steady-state response). Temperature-dependent resistive measurements[45] of PCM pixels offer a high-speed solution for understanding the average temperature of a device but require electrical contact to the PCM while offering no spatial temperature information. Time- and frequency-domain thermoreflectance techniques[46] have been utilized to perform measurements with good resolutions in both space and time, but like AFM thermoreflectance, often rely on a metallic transducer layer or probe to be a part of the system. These invasive metallic additions have the potential to affect the thermal response of the PCM (typically only a few tens of nanometers thick) and changes the optical properties of the device, limiting further characterization. As such, a new approach to measuring the thermal response of optical PCM devices is needed.

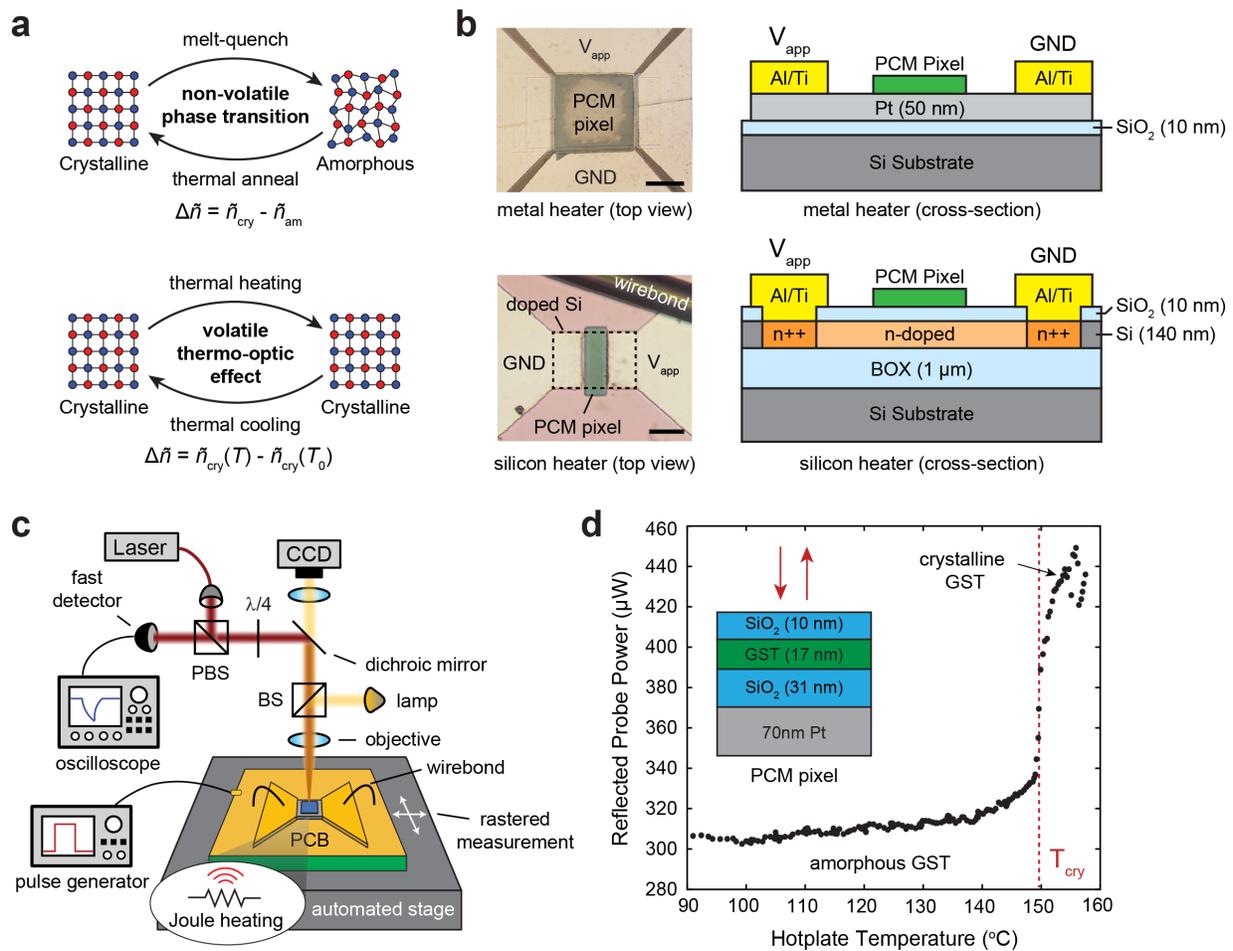

**Figure 1: Thermo-optic (TO) reflectometry in phase-change materials. (a)** Illustration of the nonvolatile phase transition (top) and volatile thermo-optic effect (bottom) in $Ge_2Sb_2Te_5$. The volatile thermo-optic effect is used to characterize the dynamic thermal response in our devices. **(b)** Microscope images and cross-sectional views of the two microheaters (Pt and doped-silicon) used in this study. Scale bars are 50 µm (top) and 25 µm (bottom). **(c)** Diagram of the reflectometry setup used to map the thermal response of metal and silicon microheaters. **(d)** Example reflectance trace of nonvolatile switching of the PCM pixel using a hotplate. The optical stack was optimized to maximize the change in reflected signal.

To address this need, we have developed a new characterization technique which leverages the volatile TO response of crystalline GST to modulate the reflection of an optical probe at normal incidence (**Figure 1a**). The TO effect is expected to be particularly strong in metavalent materials, such as GST and GeTe, due to high anharmonicity[47] and, as the dominant thermal effect within our optical stack, allows us to directly measure the thermal profile of the GST layer. In this work, we explore the thermal response of a $SiO_2$ / GST / $SiO_2$ pixel on metallic (50-nm-thick Pt) and doped-silicon (doping concentration of n++ and n-doped regions were ~$10^{20}$ cm$^{-3}$ and ~$10^{18}$ cm$^{-3}$, respectively) resistive microheaters shown in **Figure 1b**. These microheaters are similar to ones used previously to reversibly switch GSST and $Sb_2Se_3$ phase change devices[29,34] and provide a suitable platform for demonstrating our thermal characterization technique.

**Figure 1c** illustrates the experimental setup used to map the thermal dynamics of our phase change devices. A 637 nm CW diode laser (OBIS 637LX) was used as the optical probe and operated at an average power of 520 μW to avoid phase changes in the GST. For steady-state measurements shown in **Figure 1d**, the probe beam was modulated, and its reflected signal detected using a lock-in amplifier (SRS860) to increase signal-to-noise-ratio. The setup also included a heated substrate holder which was used to measure the crystallization temperature of our GST pixel as shown in **Figure 1d**.

While the TO effect can be significant for GST in both the amorphous and crystalline states[42], nonvolatile changes in the refractive index caused by incremental crystallization during electrical stimuli make characterizing the thermal response challenging. To address this, we use the volatile TO response of GST after it had been fully crystallized in our measurements. The inset of **Figure 1d** shows the optical stack of the GST pixel used. The Transfer Matrix Method (TMM) modeling approach[48] was used to optimize the thicknesses of the $SiO_2$ and GST layers within our pixel. It was found that an optical stack of 31 nm $SiO_2$, 17 nm GST, and 10 nm $SiO_2$ maximized the thermal sensitivity at the wavelength of our optical probe (λ = 637 nm) while providing sufficient encapsulation to protect the GST layer from oxidation during measurements (see **Figure 3** and following discussion for more details).

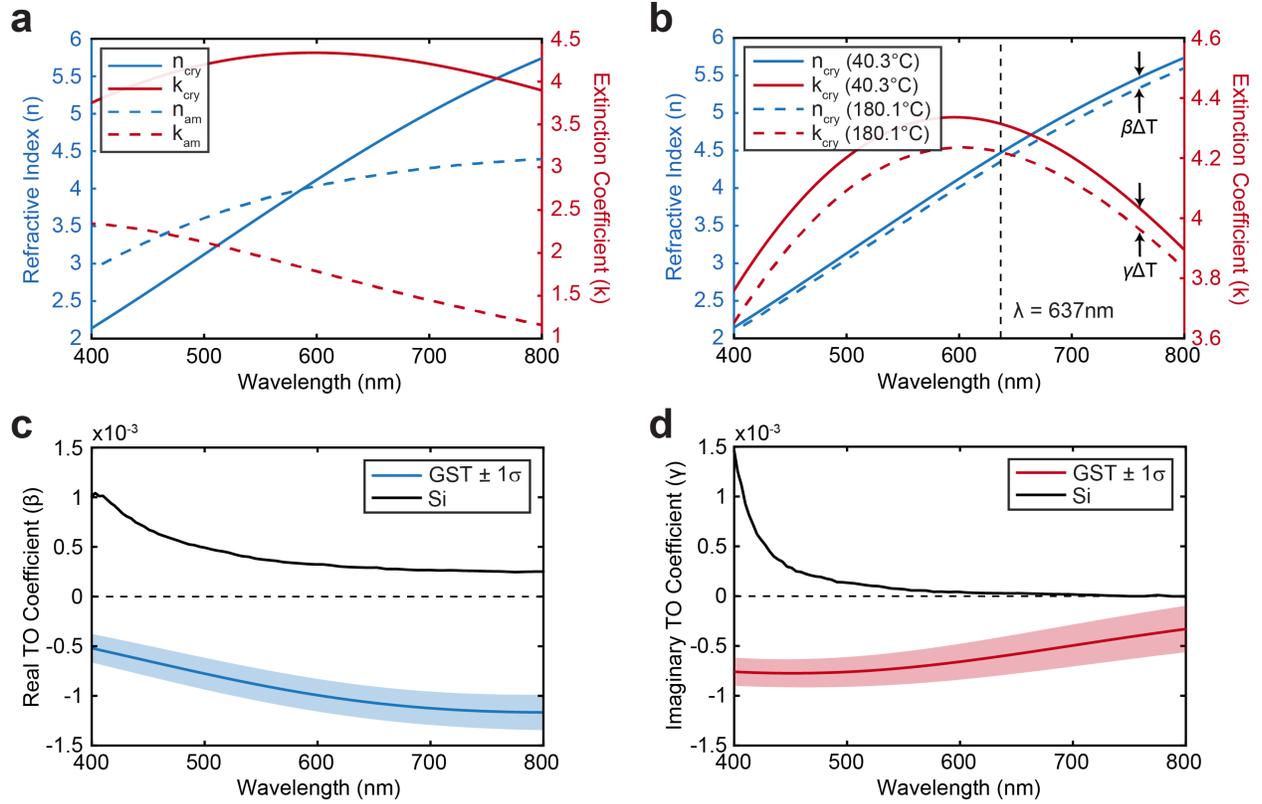

**Figure 2: Measuring the thermo-optic effect in GST thin films through temperature-dependent ellipsometry.** **(a)** Refractive index of as-deposited (amorphous) and annealed (crystalline) GST on a silicon substrate measured using ellipsometry. **(b)** Real and imaginary refractive index of crystalline GST at 40.3°C and 180.1°C. A decrease in refractive index is due to the thermo-optic response of GST. **(c)-(d)** Thermo-optic coefficients for the real ($\beta$) and imaginary ($\gamma$) components of GST extracted from temperature-dependent ellipsometry. Real and imaginary thermo-optic coefficients for silicon (solid black lines) reproduced from Vuye et. al[51].

To determine the TO response of our GST thin films, we performed temperature-dependent ellipsometry on 17-nm-thick GST sputtered on silicon substrates, encapsulated with 9.3 nm of $SiO_2$. **Figure 2a** shows the real and imaginary refractive index of GST at room temperature for both as-deposited (amorphous) and annealed (crystalline) GST extracted from ellipsometry. A significant change in both the real and imaginary components can be seen upon the phase transition which agrees well with other measurements in literature[49]. After fully crystallizing the GST (10 minutes at 250°C), we performed ellipsometry again at multiple temperatures between 40°C and 180°C at steps of 20°C using a custom-built heated stage with closed-loop temperature control. **Figure 2b** illustrates the observed change in the complex refractive index between two ellipsometry measurements at different substrate temperatures. Assuming a first-order TO effect[42], we used the following equations to model the linear change in refractive index as a function of temperature:

$$n(T) = n(T_0) + \beta(T - T_0) \quad (1)$$

$$k(T) = k(T_0) + \gamma(T - T_0) \quad (2)$$

where $n(T)$ and $k(T)$ are the temperature-dependent real and imaginary refractive indices of GST, $n(T_0)$ and $k(T_0)$ are the refractive indices at room temperature, $T$ is temperature, and $\beta$ and $\gamma$ are the real and imaginary linear TO coefficients, respectively in units of $K^{-1}$. To extract the refractive index of GST at each temperature, we used the single Tauc-Lorentz Dispersion[50] model and included the TO response of the silicon substrate in our ellipsometry models. The real and imaginary TO effect of silicon[51] is plotted as solid black lines in **Figure 2c-d**. Using equations *(1)* and *(2)*, we fit our temperature-dependent ellipsometry results at wavelengths ranging from 400–800 nm. Fits for both $\beta$ and $\gamma$ are shown in **Figure 2c-d** with the standard deviation of linear fits denoted by shaded regions. At our probe wavelength (637 nm) the TO effect of GST is clearly dominant compared to that of silicon, though we have included both the thermal response of both silicon and GST in our TMM models described below.

Using the extracted linear TO coefficients from **Figure 2c-d**, we used the TMM approach to model the temperature-dependent reflection of our GST pixels on the metal and silicon microheaters shown in **Figure 1c** at normal incidence. **Figure 3a** shows the simulated reflection spectrum of the GST pixel on top of both metal and silicon microheaters when the device is at room temperature. The 140 nm silicon device layer on 1 μm oxide gives rise to multiple reflection peaks due to optical interference compared to the reflection spectrum of the metal heater which is relatively flat. The change in the reflection spectrum as a function of temperature is shown in **Figure 3b-c** for the metal and silicon microheaters, respectively. According to our TMM model, which incorporates the TO effects of both GST and silicon, the relationship between changes in temperature and changes in reflection is linear at the probe wavelength (dashed black line in **Figure 3b-c**) for GST pixels on both microheaters. This relationship allows us to directly map changes in optical reflection of our probe to changes in temperature of the GST layer which we demonstrate in the following results. Again, the major advantage of this approach is the ability to non-invasively probe the temperature of the GST with diffraction-limited resolution and at speeds limited by the electrical bandwidth of the photodetector and readout circuitry.

To clearly illustrate the different phase-change and TO effects in GST, we performed a voltage sweep on a metallic heater with the GST pixel initially in the amorphous state. The resistance of the heater remained relatively constant and was highly repeatable for multiple sweeps (**Figure 3d**). During the forward sweep (0V to 10V), a dramatic and nonvolatile change in the reflection can be seen in **Figure 3e**, indicating a phase transition from the amorphous to crystalline state in the GST layer. However, on the return sweep (10V to 0V) and all subsequent voltage sweeps, a smaller, volatile change in reflection can be observed, indicating the TO effect in the crystalline GST layer. The inset of **Figure 3e** shows this volatile change in reflection more clearly. The reflection due to the TO effect follows a quadratic relationship with voltage since power dissipated due to Joule heating is equal to $V^2/R$, where $V$ is the applied voltage and $R$ is the resistance of the microheater. Using the reflection-temperature relationship extracted from the TMM results of **Figure 3b-c**, we can plot the temperature of the GST pixel as a function of applied electrical power. **Figure 3f-g** shows the measured temperature of GST pixels on Pt and silicon microheaters as a function of applied voltage and power density, respectively. From **Figure 3e**, the GST crystallization growth rate reaches a maximum when the metallic heater is approximately 9V. This corresponds to an applied power density of about 41 kW/cm² and temperature of 135°C which is about 15°C lower than the crystallization temperature measured on a separate device in **Figure 1d** using a temperature-controlled hotplate. While silicon and metal heaters show similar performance in **Figure 3f**, a comparison of temperature versus power density reveals that the doped-silicon heater has ~2× higher heating efficiency than the Pt microheater at steady state (**Figure 3g**).

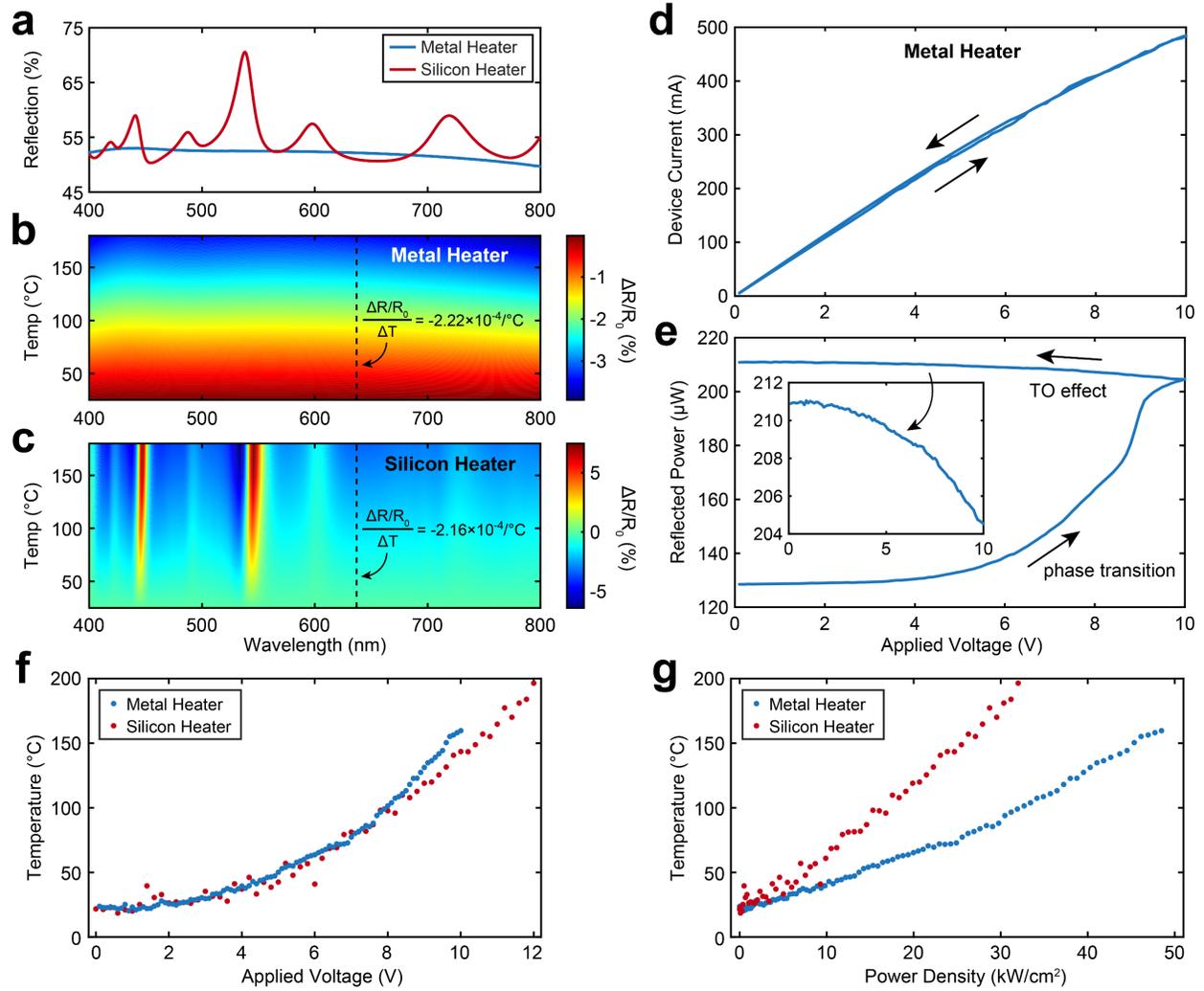

**Figure 3: Steady state thermal measurements and optical modeling of GST pixels on metal and silicon microheaters.** **(a)** Simulated optical reflection spectrum at room temperature using the Transfer Matrix Method approach. **(b)-(c)** Temperature dependent reflection spectrum for GST pixels on **(b)** metal (Pt) and **(c)** silicon microheaters (dashed lines at 640 nm denote the laser wavelength used for all reflection measurements). **(d)** Current-voltage curve for metal microheater. **(e)** Measured optical reflection of GST pixel on metal heater during the voltage sweep in **(d)**. The as-deposited amorphous GST is crystallized as the voltage increases (i.e., nonvolatile phase transition) and shows at volatile thermo-optic response as the voltage decreases (see inset). **(f)** Temperature of the GST layer as a function of applied voltage for metal and silicon microheaters (metal heaters were limited to ≤10 V to prevent permanent damage). Estimates of the temperature from changes in reflection were calculated from the temperature-dependent TMM modeling results in **(b)-(c)**. **(g)** Comparison of heating efficiency for the metal and silicon microheaters from **(f)**.

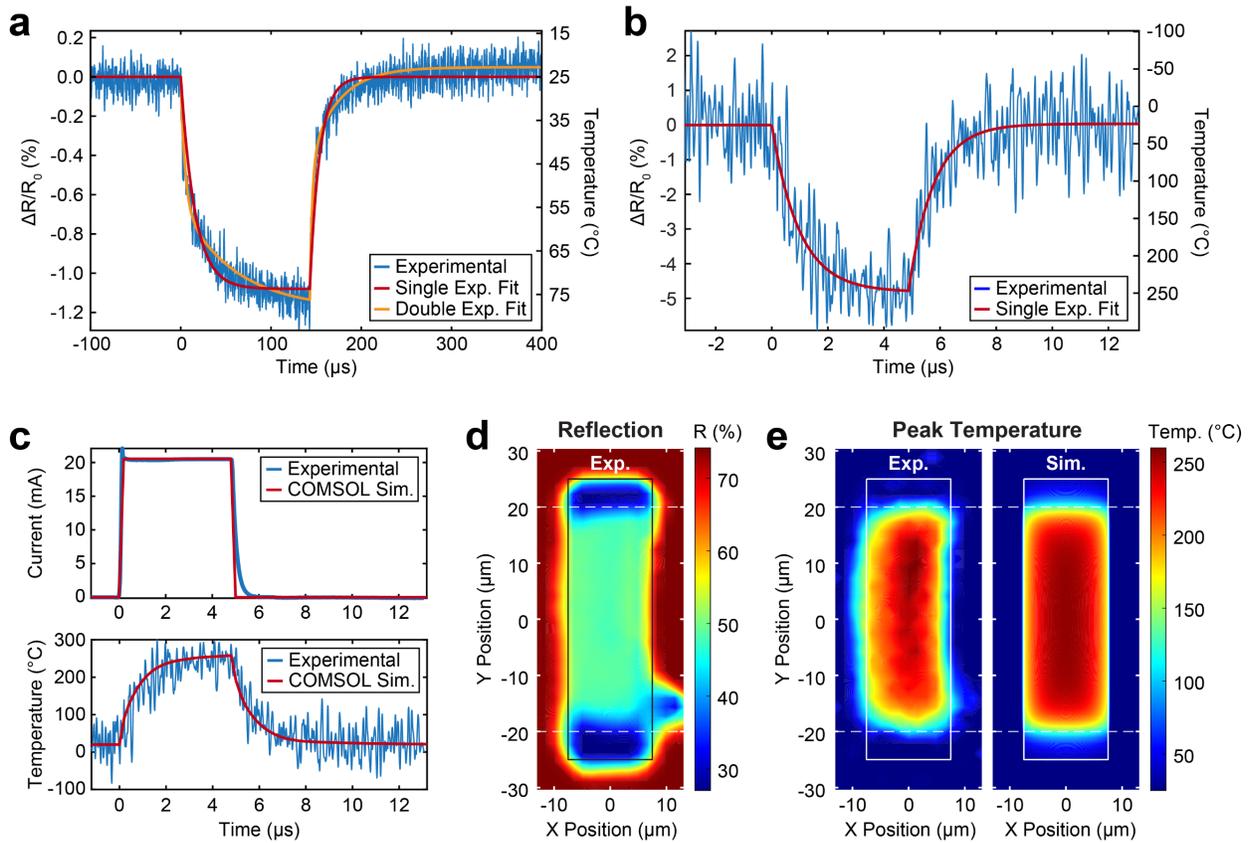

**Figure 4: Dynamic thermal response of (a) metal and (b)-(e) silicon microheaters. (a)** Dynamic thermal response of a 100×100 μm² resistive metal heater to an RF pulse (125 μs at 2.6 W) showing poor efficiency and speed. **(b)** Dynamic thermal response of a 19×40 μm² doped silicon microheater (4.8 μs at 265 mW) with improved speed and efficiency compared to **(a)**. **(c)** Time traces of simulated and experimentally measured current and temperature response of the silicon microheater, showing good agreement. **(d)** Room temperature refection map of GST (edges indicated by black outline) after crystallization from thermal annealing using the underlying silicon microheater. The GST directly above the doped-silicon heater (edges defined by dashed white line) is crystallized while the GST at the top and bottom of the pixel remain in the amorphous state. **(e)** Comparison between the experimentally measured (left) and simulated (right) peak temperature of the device during the 4.8 μs electrical pulse.

After performing steady-state thermal measurements, we turned our attention to the dynamic response of the GST pixels on the metal and silicon microheaters. For these dynamic measurements, the photodetector was connected to a transimpedance amplifier (Edmond Optics, 200 MHz bandwidth) and measured using an oscilloscope (Rigol MSO8204), allowing the thermal dynamics to be resolved with sub-10 ns temporal resolution. **Figure 4a** shows a time trace of the optical reflection for the resistive metal heater which has an active area of 100×100 µm². Using the relationship between reflectance and temperature found in **Figure 3**, we can convert the reflected probe signal to temperature. Due to the low heating efficiency of the metal heater, we used a sinusoidal RF pulse (125 µs pulse width with 70 MHz carrier frequency) and high-power RF amplifier (Mini-Circuits ZHL-5W-202-S+) to achieve an estimated 2.6 W of applied power, accounting for impedance mismatch between the amplifier and microheater. For the silicon microheater in **Figure 4b**, the device reaches a much higher temperature at much lower energies (4.8 µs pulse width at 265 mW applied power) and we were able to apply direct electrical pulses using a power MOSFET circuit similar to the approach used by Y. Zhang et al[34]. A current trace of the pulse applied to the silicon microheater can be seen in **Figure 4c**. We fit simple exponential functions to both the heating and cooling dynamics of the silicon microheaters (solid red curves in **Figure 4a-b)** using the following equation:

$$T(t) = \begin{cases} T_0 & t < 0 \\ T_0 - A\left(1 - e^{-\frac{t}{\tau_h}}\right) & 0 \leq t < t_{end} \\ \left(T_0 - A\left(1 - e^{-\frac{t_{end}}{\tau_h}}\right)\right) + A\left(1 - e^{-\frac{t-t_{end}}{\tau_c}}\right) & t \geq t_{end} \end{cases} \quad (3)$$

where $T_0$ is the temperature of the device with no pulse applied (assumed to be 25°C), $A$ is the steady state temperature bias of the device as $t \to \infty$, $t_{end}$ is the end of the electrical pulse, and $\tau_h$ and $\tau_c$ are the heating and cooling time constants of the device, respectively. From the fitting curves to the thermal response of our microheaters, we observe heating and cooling time constants of $\tau_h = 13.8 \pm 0.33$ μs and $\tau_c = 10.3 \pm 0.25$ μs for the metal heater and $\tau_h = 981 \pm 21.6$ ns and $\tau_c = 853 \pm 19.8$ ns for the silicon heater. Thus, we see that the silicon microheater design has a much faster response than the metallic heater, making it more suitable for PCMs with faster crystallization dynamics. This faster response can be mainly attributed to: (1) more efficient heating of the silicon microheater compared to the metallic one due to a much larger oxide spacing between the heater and the silicon substrate (1 μm versus 10 nm oxide spacer for the silicon and metal heaters, respectively); and (2) a ~13× smaller active heating area for the silicon heater (19×40 μm² for the silicon heater versus 100×100 μm² for the metal heater). Both an increased thermal isolation between the heater and silicon substrate and reduced heating area allows the microheater to reach a higher temperature in a shorter time, reducing the spread of thermal energy to the surrounding material. Reducing this parasitic heating of the substrate also reduces the heated volume of the system, enabling faster quenching times. This can also explain the observed deviation of the metallic response from a single exponential function. From **Figure 4a**, it appears that two heating and cooling time constants are at play due to non-negligible heating of the substrate. This effect has been observed before in graphene thermal emitters and PCM microheaters where the in-plane versus out-of-plane heating and cooling rates of the microheater and substrate differ[38,52]. To demonstrate this effect, we fit a double exponential function (solid orange line) to the thermal response in **Figure 4a** and observe both fast heating and cooling time constants of the metal heater ($\tau_h = 4.64 \pm 0.30$ μs and $\tau_c = 2.76 \pm 0.23$ μs) as well as slow heating and cooling time constants of the silicon substrate ($\tau_h = 60.0 \pm 2.26$ μs and $\tau_c = 31.2 \pm 1.77$ μs) which differ by an order of magnitude.

Due to the low heating efficiency of the metal heater, we limited our attention to the silicon microheater and used COMSOL Multiphysics®[53] to simulate its thermal response during an applied current pulse. A comparison between experimental and simulated results can be seen in **Figure 4c**. The model uses the electric currents module coupled with heat transfer in solids module to simulate Joule heating. To ensure the applied electrical power in the simulation properly matched our device, the electrical conductivity of the doped silicon layer was derived from the device's measured current-voltage response and imported into COMSOL as a function of applied voltage. The rest of the material properties used can be found in **Table 1**, those properties marked as 'n/a' are due to the fact that the electric current module was only applied to the active area of the thin film Si and metal contacts for model simplicity.

| Material | Electrical Cond. $\sigma$ [S/m] | Thermal Cond. $k$ [W/(m·K)] | Specific Heat $C_p$ [J/(kg·K)] | Density $\rho$ [kg/m$^3$] |
|---|---|---|---|---|
| Si (thin film) | from IV | From [54] | From [55] | 2329 |
| Si (bulk) | n/a | 130 | 700 | 2329 |
| SiO$_2$ | n/a | 1.4 | 730 | 2200 |
| Al | $3.776 \times 10^7$ | 238 | 900 | 2700 |
| GST | n/a | 0.19[20] | 213[20] | 5870[20] |

**Table 1:** Material properties used for COMSOL simulations of the doped-silicon microheater.

In addition, a thermal contact resistance of $2 \times 10^{-9}$ K·m$^2$/W, $7.69 \times 10^{-9}$ K·m$^2$/W, and $2 \times 10^{-9}$ K·m$^2$/W were used for the Si/SiO$_2$, Si/Al, and GST/SiO$_2$ boundaries, respectively[41]. Both the bottom surface of the Si chip and the top surfaces of the Al contacts are held at constant room temperature of 25°C. The Al metal contacts were also modeled as heat sinks due to their excellent ability of conducting heat, as well as the wire bond's ability to conduct the excess heat away from the contacts. We see excellent agreement between the measured and simulated thermal traces at the center of our device (shown in the lower panel of **Figure 4c**), indicating that our COMSOL model captures the thermal response of our silicon microheater.

We also compare the experimental and simulated spatial thermal profile of our device in **Figure 4d-e**. **Figure 4d** shows the optical reflection of the GST pixel (boundaries indicated by solid black lines) on top of the silicon microheater (boundaries of doped silicon region indicated by dashed white lines). After thermal annealing of the GST layer by applying multiple 0-10 V sweeps, we can observe a clear contrast between the crystallized GST directly on top of the microheater (green solid area) and the amorphous GST covering the undoped silicon (blue areas). This indicates that the heating is highly localized in the doped silicon region as expected. In **Figure 4e**, we compare both the experimental and simulated thermal profile of the GST pixel at the peak temperature which coincides with the end of the electrical pulse. This is achieved by rastering the device under the probe beam while recording the thermal response at each spatial position (illustrated in **Figure 1b**). While the temperature in the center of the pixel is in good agreement with our COMSOL simulation, we also see some slight deviation between experiment and theory, especially near the corners of the pixel, which could indicate non-idealities during device fabrication. This highlights the usefulness of having an experimental technique to probe the fabricated device, rather than purely relying on simulations.

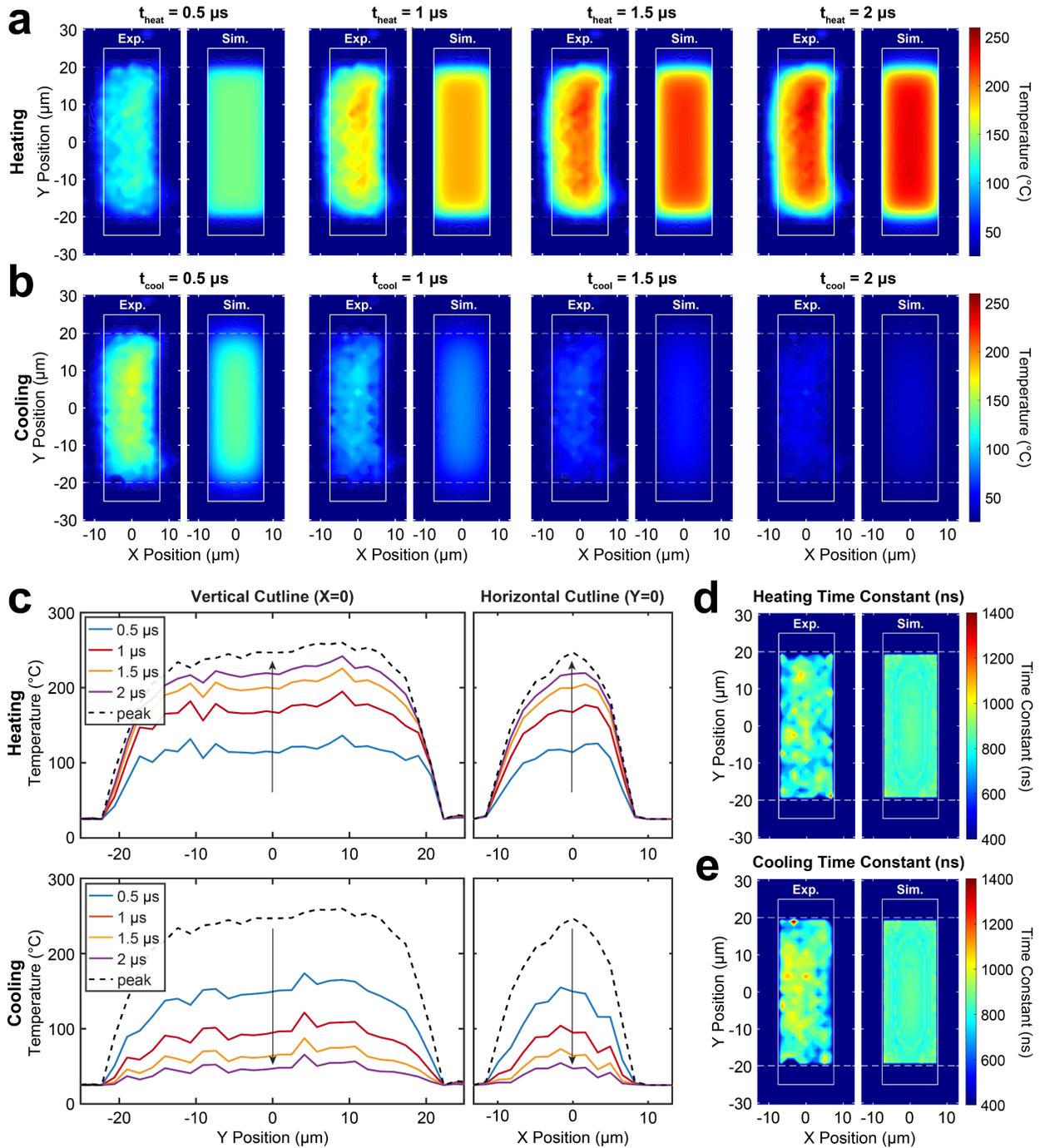

**Figure 5: Spatially mapping the dynamic thermal profile of a silicon microheater. (a)-(b)** Dynamic thermal profile at 500 ns time steps during **(a)** heating and **(b)** cooling of a doped silicon microheater. Experimental results of thermal profile (left) agree well with our simulations (right). **(c)** Vertical and horizontal cross-sectional cuts of the heating and cooling profiles from **(a)** and **(b)**. **(d)** Heating and **(e)** cooling time constants extracted from exponential fits to the dynamic thermal response at different positions in the device. Both experimental (left) and simulation (right) show that the time constants are spatially independent within the GST layer. Variations in the experimental data are attributed to uncertainties in the exponential fit.

**Figure 5a-b** compare the experimental and simulated thermal profile across the device during the first 2 µs of both the heating and cooling process. We define $t_{heat}$ as the time measured from the start of the applied electrical pulse and $t_{cool}$ as the time measured from the pulse end. Again, we see excellent agreement between experiment (left) and simulation (right). Cross-sectional cutlines across the vertical and horizontal centers of the device are shown in **Figure 5c**. We see that the heating profile of the device is fairly uniform and we are limited by the resolution of the probe beam (FWHM of 4.7 µm) close to the edges of the pixel.

We can also extract the heating and cooling time constants as a function of position across our GST pixel (**Figure 5d-e**). As both the melting temperature and quenching rate of the PCM determines whether or not it can be re-amorphized and thus reversibly switched, significant spatial differences in either the peak temperature or cooling time constants can lead to a device which is able to reversibly switch only a portion of the total PCM area. **Figure 5d-e** show the extracted heating and cooling time constants across the silicon heater area using the exponential fitting equation *(3)*. While there is variation across the experimental time constants, it appears to be random and can be attributed to the quality of the fit. The heating and cooling time constants averaged across the GST pixel were found to be $\tau_h = 815 \pm 139$ ns and $\tau_c = 843 \pm 159$ ns, respectively. This agrees very well with fits to our COMSOL simulation which yielded $\tau_h = 823 \pm 45.3$ ns and $\tau_c = 843 \pm 35.3$ ns when averaged across the GST pixel.

**Conclusion:**

In summary, we have developed a simple yet powerful technique to non-invasively probe the local temperature of phase-change devices by leveraging the strong TO effect in GST. We used this technique to investigate two electro-thermal designs that have been used previously by the phase-change community and directly compare their relative performance. This enabled us to determine crucial metrics for electrically-programmable PCMs, such as heating efficiency, speed, quenching rate, and spatial uniformity. For the silicon microheater, our experimental results agreed well with modeling results near the center of the device but highlighted the need for experimental characterization of actual devices after fabrication. We anticipate that the application of our technique will provide useful insights into the design and optimization of robust and reversible phase-change devices which are electrically controlled, paving the way to large-scale integration.

**Author Contributions:**

N.Y., F.X., J.H., and C.R. conceived the experiment. S.V., C.R. and Y.Z. fabricated the microheaters. N.N. deposited the $Ge_2Sb_2Te_5$ pixels, implemented the measurement setup, and performed all experiments. J.E. performed COMSOL simulations of the devices. All authors discussed the data and wrote the manuscript together.

**Funding Sources:**

This work was supported in part by the U.S. National Science Foundation under Grants ECCS-2028624, ECCS-2210168/2210169, DMR-2003325, ECCS-2132929, and CISE-2105972. N.Y. acknowledges support from the University of Pittsburgh Momentum Fund. C.R acknowledges support from the Minta Martin Foundation through the University of Maryland. Approved for public release. Distribution is unlimited. This material is based upon work supported by the Under Secretary of Defense for Research and Engineering under Air Force Contract No. FA8702-15-D-0001. Any opinions, findings, conclusions, or recommendations expressed in this material are those of the author(s) and do not necessarily reflect the views of the Under Secretary of Defense for Research and Engineering.